# MSS: a new way for high energy-resolution on any accelerator when using magnetic spectrometers

A. Gafarov[1*]

[1]Nuclear Reactions Laboratory, INP of Science Academy of Republic Uzbekistan

[*]email: anatxagor@gmail.com

An innovative approach: **MSS** - the ***Method of Spectra Superposition*** for high energy-resolution (*ΔE)* studies on any accelerator of nuclear reactions at fixed targets by magnetic spectrometers, has been invented. For **thin** targets **MSS** provides a *ΔE* which depends only on the detector's energy-resolution.

**The MSS can bring huge benefits to experiments with fixed targets at all types of accelerators and also on colliders.**

The **MSS** was used to get precise data with a *ΔE* =10 keV in $p + {}^{12}C$ elastic scattering for the energy region $E_p$ 16→19.5 MeV at cyclotron U-150 of **INP** Ulugbek (Tashkent, AS of Uzbekistan) with a beam energy spread of ∼200 keV. The proton energy control was performed with a 20-step Energy Moderator which provides $E_p$ interval of 3.5 MeV without any energy-readjustment of cyclotron and all the ion optics of a 41-meter-long beam pipe. A multichannel Magnetic Spectrograph (**Apelsin**) was used as the nuclear reaction products detector. The particle-products were detected on focal planes of **Apelsin** by specially designed Coordinate-sensitive gas MWPCs with two particle counters in coincidence. The acquisition system was based on IBM PC online with a fast CAMAC branch custom-designed at PNPI (St-Petersburg INP, USSR) together with fast TDC modules and other fast electronics providing immediate selection & accumulation of events from the MWPCs. The **Apelsin** magnetic field was stabilized to 3 ppm by NMR-monitor system The obtained, with statistical error ∼6%, excitation function (**EF**) of ($p, {}^{12}C$), with a resonance reach structure, is in precise agreement with thresholds and levels data [15-18].

## 1. Method of Spectra Superposition: Basics

Historically, particle detectors based on magnetic field have proved most important when the goal is to quickly analyze large numbers of events: with separation of components by momentum (energy), electric charge and mass – and all with the highest accuracy. The magnetic field has no limitations like incoming flux intensity and radiation hardness. But a question often arises: **why ask for high energy-resolution detection if the accelerator's energy-spread will wash it away?** Although the combination of a cyclotron and a high energy-resolution magnetic spectrograph may at first seem strange if not silly, the **MSS** provides a good counter example [1-6]. Let's see why. **Obviously, each single beam-particle** could be considered as a Dirac **δ-function on the energy scale.** If one could maintain this delta-function through the whole measurement process, that would suffice for highly accurate data. So, the accelerator's beam energy-spread itself is not an obstacle.

But there are some rules for experiments:

1) The **Beam intensity** has to be sufficiently low that the rate for pile up of interactions within the target particles is negligible.

2) Distortions to the initial beam-particle's delta function energy spread, introduced by straggling of beam particles in the **target, should be** significantly smaller than the required final energy resolution, *ΔE* ).

Without energy-tagging of initial projectiles, **one can measure only energy-dependencies for reactions observed.**

Even in a **thin target,** the *δ*-energy-spectrum of each projectile is distorted. As shown in Fig. 1, the spectrum is broadened and shifted to lower energies. The distortions, in a realistic energy distribution, are represented by a superposition of Gaussian peaks.

The projectile scattering by the target electron "gas" results in a shape looking like a mixture of both the Landau and Blunck-Leisegung distributions (spectrum on beam-energy scale) [7-9].

The target must be **thinner** than the one that causes losses (the gray band) of total $\int \delta = 1$ area more than 1% (the main Gaussian term, must contain $\geqq 99\%$). This shape represents probability, for the single projectile-particle, to lose initial energy and be moved from the $\delta$-spectrum to the left by straggling with the target electrons.

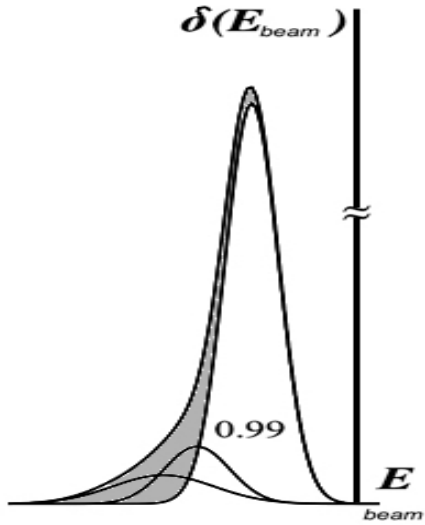

Fig.1.

Combination of cyclotron and a magnetic spectrograph paves straight way for measurements of the energy-dependent $\sigma(E)$, or/and $d\sigma(E)/d\Omega$.

The Fig. 2 shows how things evolve in the **MSS**: initial $d\sigma_A(E)/d\Omega_\theta$ shape of some process **A** is shown on the top, then the accelerator's beam initial energy spectrum enters the target body, where gets a little wider and shifted $J(E_n)$ to the lower energies immediately before interaction, then the particle-product spectrum - $I_A(E_p)$ leaves interaction point, passes through the rest of target and gets a little washed and shifted to the lower energies. Next below is the detector's Response Line the $R(E,E_f)$ which then, in convolution with $I_A(E_p)$, forms the final energy spectrum $S_A(E)$ of detected products of process **A**.

More carefully about all on the Fig. 2 can be described mathematically as a series of transforms.
The projectile-particles of accelerator before the interaction (of type **A**) pass through the target body with electrons **e** that cause energy-losses (Fig. 1, no transverse momenta of **e**-struggling accounted) what results in an expression (1):

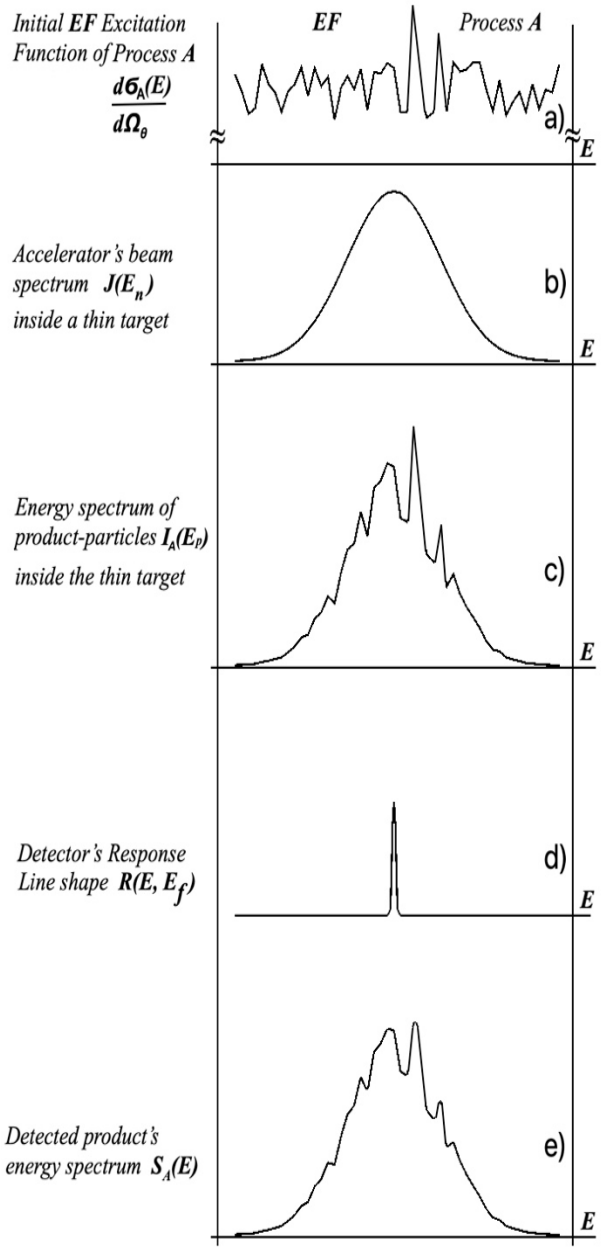

Fig. 2. **(a)** – Energy dependence of differential cross-section $d\sigma_A(E)/d\Omega$ of "process **A**" on the top and its transformation step-by-step when bombarded with the **(b)** - accelerator beam of shape $J(E_n)$ at the **thin target** body - right before interacting with the target-particle, **(c)** - the process **A** sharp energy-spectrum of the product-particles immediately after interaction or behind the **thin target**, **(d)** - detector's Response Line $R(E, E_f)$ for the process **A** product-particles with energies $E_f$, **(e)** - the detected product-particle spectrum $S_A(E)$.

$$J(E_n) = \frac{J_o}{\sqrt{2\pi(\sigma_o^2 + \sigma_\varepsilon^2)}} \sum_o^s c_s \, exp\left\{-\frac{[E_n-(E_o-\varepsilon_s)]^2}{2(\sigma_o^2+\sigma_\varepsilon^2)}\right\} \quad (1),$$

where $J_o$, $E_o$, $\sigma_o$ - the initial accelerator's beam intensity, central energy (median) and standard deviation from $E_o$ respectively, $c_s$ –coefficients of expansion of the distorted beam particle $\delta$-spectrum (grey + white on Fig. 1) into a series $s$ of Gaussian peaks, $\varepsilon_s$ – the centroid of Gaussian "term-s" in the expansion. $E_n$ –energy of projectile-particle right before the interaction (process $A$), $\sigma_\varepsilon$ –standard energy-loss deviation from $\varepsilon_s$ in the Gaussian term #$s$ (Fig.1). As one can see, electronic straggling causes some widening of the accelerator's initial energy-spectrum and, taking into account only first Gaussian term (99%) from Fig.1, would provide enough accurate description of the beam energy-spectrum change in the thin target (right before the interaction).

Now let's suppose that, our target will show interactions processes ($A,B,C,...$) which will never overlap because of the accelerator's energy spread (the g.s. and other excited states of target-particles are well separated on the energy axis).

Without the ionization energy-losses for the products-particle of process $A$ in the target medium (or in case of ideally thin target) the product-particles energy-spectrum would be pure excitation function $d\sigma_A(E)/d\Omega$ of process $A$ piloting the beam shape $J(E_n)$ what gives $I_A(E_p)$ on the product energy axis. Making an expansion like shown on Fig.1 but for the product-particles (of $A$) one will have a series of $r$ Gaussian terms with $\sigma_\xi$ - standard deviation and magnitudes $m_r$.

Let's take that in this small energy range of $\pm 3\sigma_\xi$ the beam intensity $J(E_n) \approx$ **const**, in this case the products energy-spectrum from target looks like this:

$$S_A(E_f) \cong \sum_o^r m_r \int_{-3\sigma_\xi}^{3\sigma_\xi} \frac{d\sigma}{d\Omega_\Theta} e^{\left\{-\frac{[E_p-E_f-\xi_r]^2}{2\sigma_\xi^2}\right\}} dE_n \quad (2),$$

where $E_f$ is energy of the product-particle slightly "slowed down" in the target medium, $\xi_r$ is the median of energy-losses expansion $r$-term.

As one can see from (2) the ionization losses is first "non-nuclear" source of distortion that slightly washes out the initial $d\sigma_A(E)/d\Omega$ of process $A$.

Let's take the Detector reproduces the product-particle energy $E_f$ as $E_D = \eta E_f$ ($\eta = $ **const**) and its Response Line is a normal distribution $R(E,E_f)$ (Fig2.d) with a standard deviation $\sigma_D$. In this case the detected spectrum is a convolution (3) below:

$$S_A(E) = \int_{-\infty}^{\infty} S_A(E_f) \, R(E,E_f) \, dE_f \quad (3),$$

or in full form it becomes (4):

$$S_A(E) \cong \sum_o^r J_r \int_{-3\sigma_\xi}^{3\sigma_\xi} \frac{d\sigma_A(E_n)}{d\Omega_\Theta} e^{\left\{-\frac{L[E_n-Q_A-\Delta E_k-E_f-\xi_r]^2}{2(\sigma_D^2+\sigma_\xi^2)}\right\}} dE_n, (4),$$

where $J_r = J(E_n) \frac{m_r}{\sigma_D \sigma_\xi} \sqrt{\frac{\alpha^2+\alpha/\varrho+\varrho+1}{8\pi(\sigma_D^2+\sigma_\xi^2)}}$, and also

$$\alpha = \frac{E}{E_f} - \eta; \quad \varrho = \sigma_D^2/\sigma_\xi^2; \quad L = 4(\varrho+1)^2 + \varrho + 1;$$

Last expression describes in general the "resonant" and background components of the energy-dependent cross-section and it's valid for **all processes** initiated by this kind of hadron-projectiles at the nuclear target. It clearly shows that the energy-spectrum of projectile-particles is just a statistical base "feeding" display of the $d\sigma_A(E)/d\Omega$ of prosecc $A$ (Fig. 2, e).

Also (4) says that shape (e) get distorted by both:
**a)** product-particle ionization losses on their path and
**b)** detector's energy-resolution (Response Line FWHM).

Influence of **a)** could be minimized by decreasing the target thickness and rising the vacuum quality what makes the detector's quality **b)** the main source of EF distortions.

The excitation function appears on the wide top of the beam energy-spectrum, so its wideness – such a non-monochromaticity becomes a positive and very useful factor in case of the EF-measurements. The only limitation is when width of beam energy-spread causes overlapping of different processes pronounced in the ordinary spectrum of products for given nuclear target (for example among the elastic and inelastic peaks).

Now it's very important to provide two condition:
- **1** small-thickness targets allowing to take into account only main Gaussian-term (99%) of the product-particle losses expansion (Fig. 1-similar) and
- **2** **flat top** of the accelerator's beam energy-shape
  $J(E_n) = $ const $= J_b$ for $E_n\{E_{0n} \pm \Delta E_{beam}/2\}$.

It this case the measured EF gets equal statistical support in each energy bin $dE_{beam}$ in the energy range $\Delta E_{beam}$ on the top, so there is no need to deal with complicating integrals. The **MSS** is based on both and expression (4) becomes a strong equality (5):

$$S_A(E) \equiv J_b \int_{-3\sigma_\xi}^{3\sigma_\xi} \frac{d\sigma_A(E_n)}{d\Omega_\Theta} e^{\left\{-\frac{L[E_n - Q_A - \Delta E_k - E_f - \xi_r]^2}{2(\sigma_D^2 + \sigma_\xi^2)}\right\}} dE_n, \quad (5),$$

where $J_b =$ **const** for $E_n\{E_{0n} \pm \frac{\Delta E_{beam}}{2}\}$, with $E_{0n}$ -the centroid beam-energy from accelerator.

Condition **1** is obvious, but what about **2**?

## 2. Energy-moderator

The answer is in shift of projectile-particles energy step by step in a special way that builds a resulting flat-top beam-energy spectrum.

If one have the beam energy-spectrum in a Gaussian shape with median $E_o$ and FWHM=$\Delta E_o$, then its standard deviation is $\sigma_o = \Delta E_o / 2.36$

How close have to be the neighboring Gaussians to start forming flat top?

The needed intermediate distance is $\Delta\varepsilon = \sigma\sqrt{3}$ (6)

Easiest way to change accelerator's energy is to use **E**-moderators –energy change in seconds, no needs in readjustment of all accelerator with a lot of the following ionic optics on the beam-pipe delivering projectile-particles to experimental setup.

Using a set of **Al**-foils before the target one can slow down the beam-particles, but let's remember that, the thicker is **Al**-plate – the wider are both - the beam solid angle-divergence and energy-spread. And in addition, the collimating system in front of target will cut away the more of beam, the thicker is **E**-moderator's plate. Because of the beam energy-spread rise, condition (6) changes to (7): $E_i - E_{i+1} = \Delta\varepsilon_i = \sigma_i\sqrt{3}$ , (7)

Calculations [1-6] of the **E**-moderator are in the Table 1 below. 19 plates (20 steps) cover an energy-range of 3.5 MeV of cyclotron U-150 **p**-beam of initial $E_{0p}$=18 MeV. Such a special **MSS**-moderator was designed in form of 20-segments ring with a window and 19 $^{27}Al$-plates of different thickness $h_i$ from **Table 1**. $\sigma_i = \sqrt[2]{\sigma_o^2 + \lambda_i^2}$ ;

| i | $E_i$ MeV | $\Delta\varepsilon_i$ MeV | $\sigma_i$ MeV | $\lambda_i$ MeV | $h_i$ μM | $K_i$ times |
|---|---|---|---|---|---|---|
| 0 | 18.000 | 0.0 | .062 | .0 | 0 | 1.0000 |
| 1 | 17.893 | 0.107 | .065 | .0195 | 20.3 | 0.9158 |
| 2 | 17.780 | 0.220 | .069 | .0293 | 39.8 | 0.9453 |
| 3 | 17.661 | 0.339 | .072 | .036 | 60.3 | 0.984 |
| 4 | 17.530 | 0.463 | .075 | .043 | 82.0 | 1.0341 |
| 5 | 17.416 | 0.594 | .079 | .049 | 104.8 | 1.0843 |
| 6 | 17.269 | 0.731 | .083 | .0545 | 128.5 | 1.1341 |
| 7 | 17.126 | 0.874 | .086 | .0597 | 153.3 | 1.1822 |
| 8 | 16.977 | 1.023 | .089 | .0643 | 179.2 | 1.2469 |
| 9 | 16.822 | 1.178 | .091 | .067 | 206.0 | 1.2389 |
| 10 | 16.664 | 1.336 | .096 | .0734 | 233.5 | 1.3267 |
| 11 | 16.498 | 1.502 | .1 | .0785 | 262.4 | 1.3749 |
| 12 | 16.325 | 1.675 | .103 | .0822 | 292.4 | 1.4195 |
| 13 | 16.147 | 1.853 | .107 | .0866 | 324.2 | 1.4686 |
| 14 | 15.963 | 2.037 | .110 | .091 | 355.2 | 1.5217 |
| 15 | 15.772 | 2.228 | .113 | .095 | 388.3 | 1.5699 |
| 16 | 15.576 | 2.424 | .117 | .099 | 422.3 | 1.6048 |
| 17 | 15.373 | 2.627 | .121 | .1035 | 457.5 | 1.6931 |
| 18 | 15.164 | 2.836 | .124 | .1075 | 493.7 | 1.8421 |
| 19 | 14.949 | 3.051 | .127 | .111 | 531.0 | 2.1926 |

## 3. The MSS-Algorithm

Exposition term $K_i$ reflects beam-charge integrated by the Faraday cup during exposition number $i$ {0,… ,19}.

Every exposition produces a typical products-spectrum (grey filled) as shown on Fig. 3 with elastic and inelastic peaks displaying pronounced fractions of the pilot-EF of processes **A, B, C,** on the top of beam energy-spectrum.

For a full cycle of 20 steps, if one combines together all 20 beam peaks on one axis, it gives a **trapezoid flat-top** beam energy-spectrum - see Fig. 4 (Q pink) below. After each exposition one have corresponding spectrum of product-particles at the set (R) (with peaks of elastic -$A_i$ and also inelastic processes $B_i$, $C_i$, $H_i$).

After data preparation by the "Spectra Superposition" procedure, one have set of ruffly pronounced, almost pure excitation functions $d\sigma(E)/d\Omega_\Theta$ of reactions A, B, C and H.

Further data processing will give the high energy-resolution EF for all observed and statistically enough supplied interactions within set (S) on Fig. 4.

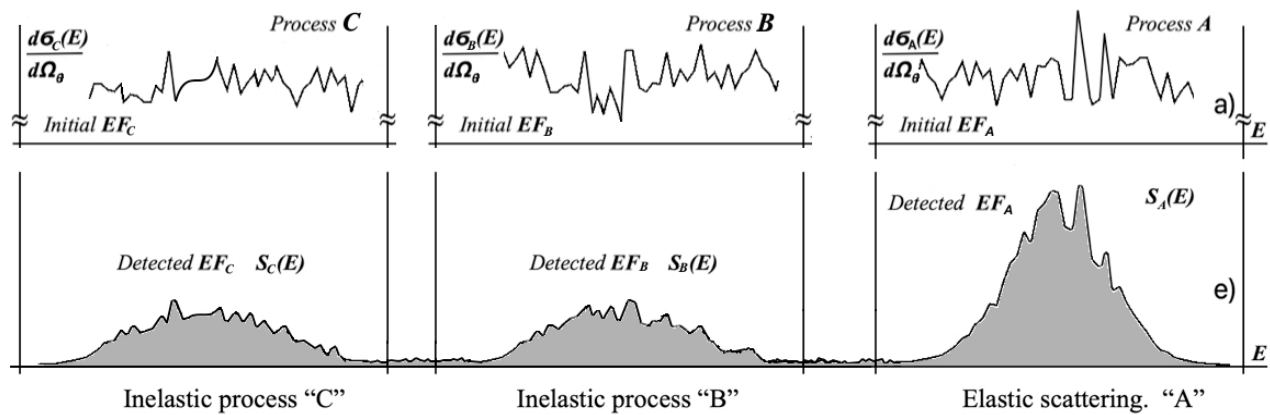

Fig. 3. Excitation functions of interactions **A, B,** and **C** with typical HR-detected spectrum of product-particles.

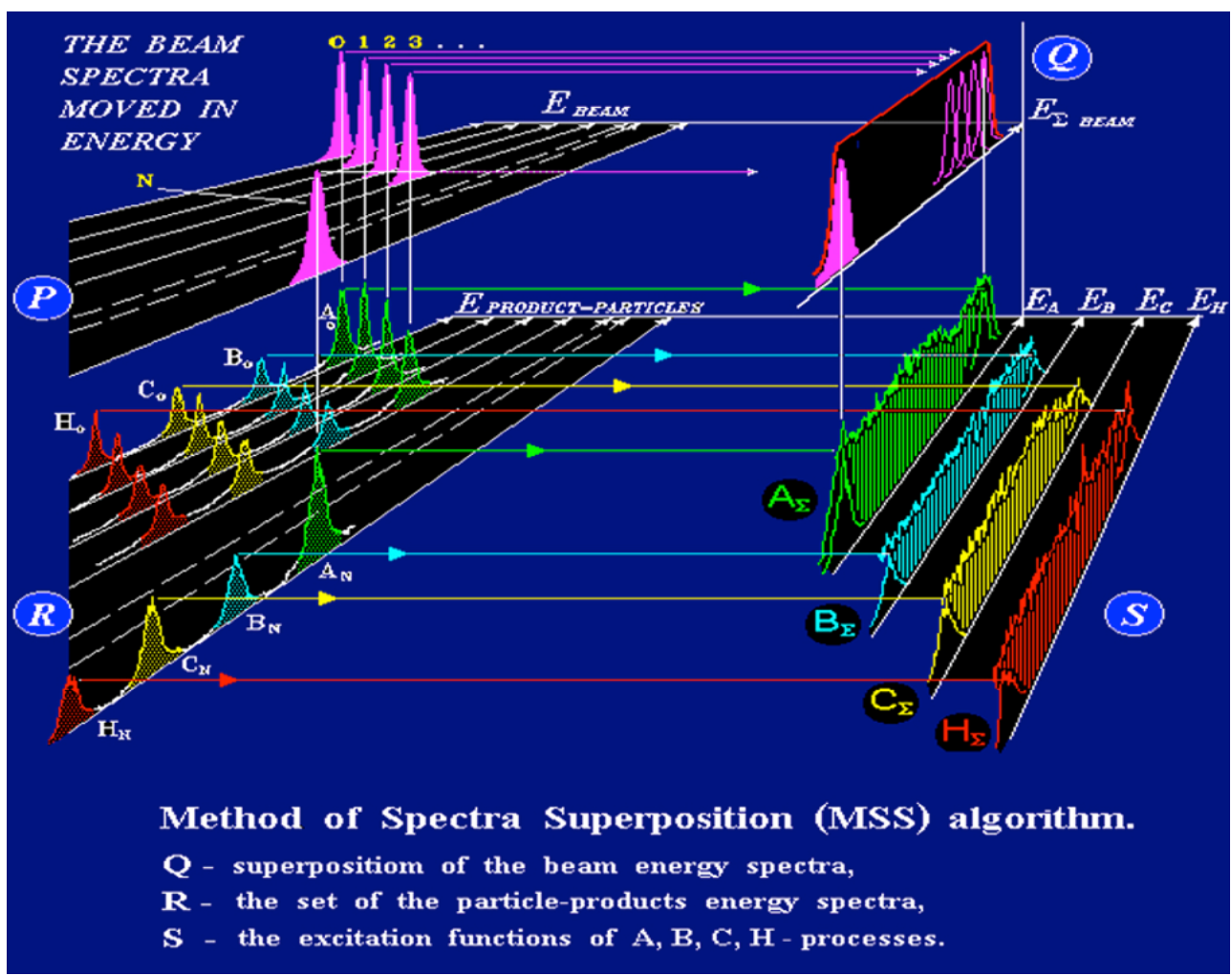

Fig. 4. The **MSS**-algorithm: beam-particle peaks (**P**) and beam superposition spectra (**Q**), interaction peaks (**R**) ($A_i$, $B_i$, $C_i$, $H_i$) with superposition set (**S**) of the elastic scattering **A** and inelastic reactions **B**, **C**, and **H**.

## 4. Experimental MSS-setup

The setup on Fig. 5, included a cyclotron U-150 with $E_p \leq 22$ MeV, the 41-meter long beam pipe with 3 magnetic prisms, 10 quadrupole lenses, and also the 14-angular Magnetic Spectrograph "Apelsin" as product-particle detector. The FWHM of "Apelsin" is better than 5 keV for protons (15÷20 MeV) [1].

The "Apelsin" magnetic field was stabilized to 3ppm by NMR-monitoring system.

To get the real time data from focal planes of "Apelsin", a special system of coordinate gas MWPC was designed together with the St-Petersburg PNPI (former Leningrad INP), light green on Fig. 6 below.

The MWPC have in one gas "pool" two coordinate proportional counters with *Au-*covered wolfram anode wires and coordinate sensitive flat cathode "rulers" with 300 ns delay lines connected to **PA** on both ends. Both counters in a MWPC work in coincidence (fast signals from both anodes and signals from both cathode delay lines). The chamber of 8*Ar* +*Xe* +*CO$_2$* mix gas was separated from the vacuum space with 7$\mu_M$ aluminized mylar film. Under the atmosphere pressure the gas mixture was slowly non-stop flowing from the high-pressure gasholder through the MWPC volume then back into atmosphere during all beam time [1,10-13].

The coincidences allowed to control angle of the incoming detected product-particles and insure really deep suppression of any background radiation.

The anode wire of the leading "upper" counter was embedded into the focal plane of section #7 of the "Apelsin" (under angle of 16.3° to the beam pointer).

Fast signals from both ends of the cathodes - flat delay-lines were preamplified [10] right there by **PA** by special schematics on fast FETs, that provided a total time-jitter ≤ 100 ps. The anodes time-jitter was ≤ 10 ps.

To provide the deepest background radiation suppression, a module [FD]+PMT-based fast-plastic scintillator counter was introduced as monitor of the real-time beam micro-bunches, via the $\gamma$-bursts from the luminophore-covered gate-slit at the reaction-chamber entrance of "Apelsin" (the dark blue filled area -vacuum). For whole acquisition system, the U-150 *HF*-generator's signal via [BS] and [FD]+PMT chain were the basic in production of major trigger.

As shown on Fig. 6, product-particles from target at the reaction chamber center, fly away through entrance slits into analyzing sectors. Magnetic field in gaps of each two sections of the toroidal core (yoke gaps) transforms the product-beam into the angle-distributed spectrum focused on the anode wire of upper (leading) coordinate counter of MWPC.

Delay time difference in signals from both ends of MWPC-cathodes twice marks trajectory-coordinates of each registered particle with a jitter of ≤ 0.1 mm on the focus plane and behind it to select right entrée angles.

The acquisition system was based on IBM PC that controlled a CAMAC-branch with special set of the nanosecond-fast modules designed at PNPI (St-Petersburg INP, USSR) including fast TDC and other fast electronics.

Such an architecture provided immediate selection & collection of events from MWPCs and Faraday cup with the [I/F]-convertor of the beam-current into frequency.

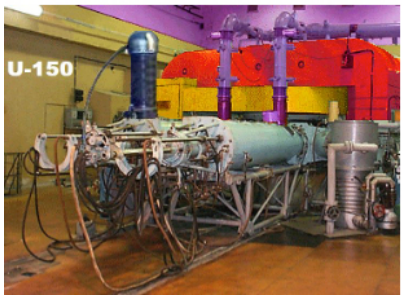
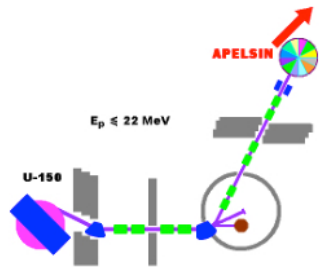
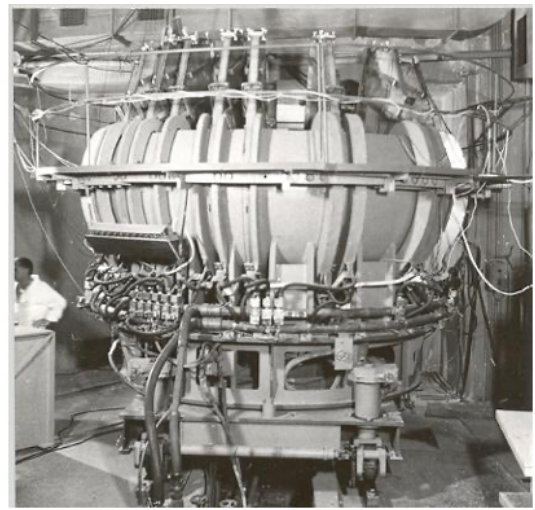

Fig 5.     Cyclotron U-150.          Sketch of experimental area.      14-angular Magnetic Spectrograph "Apelsin"

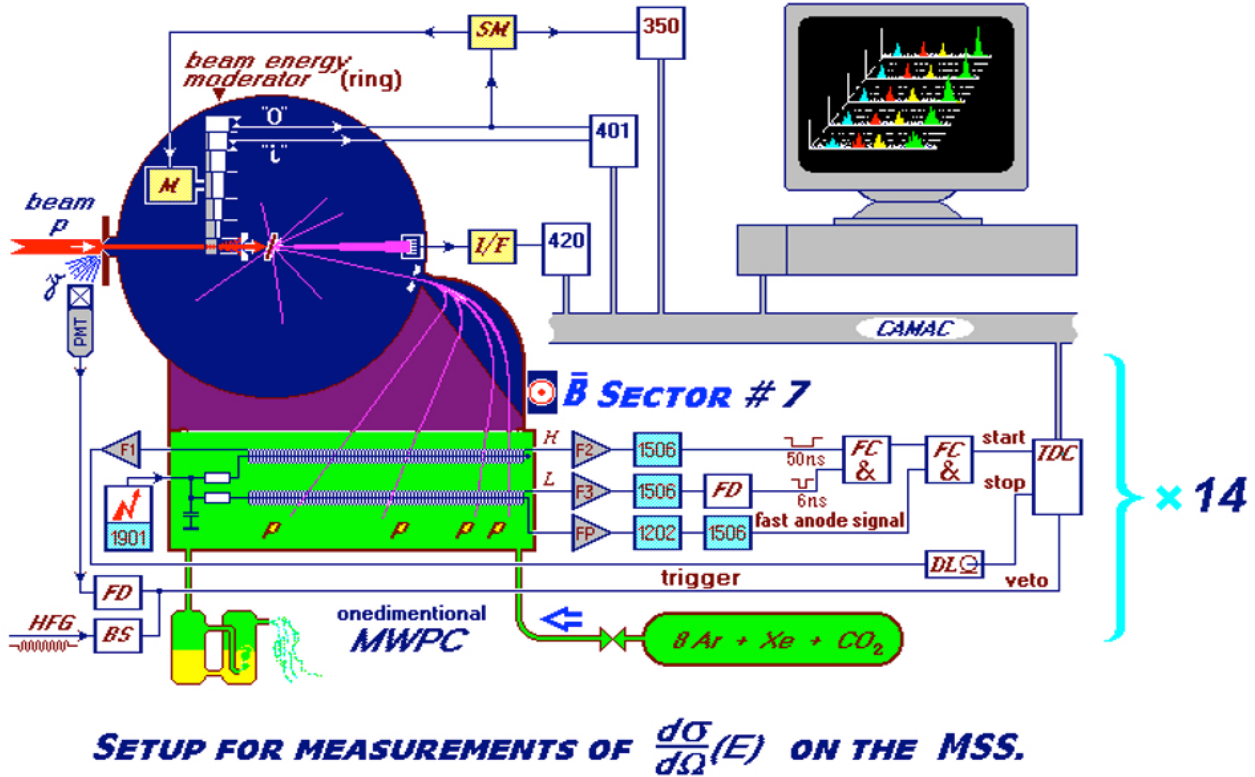

Fig. 6

## 5. The MSS-measurement procedure

The **MSS**-setup works in cycles of 20 expositions of target under 20 different beam-energies.

First of all, one chose the initial beam-energy $E_o$ what means one will have an energy range of ~ 3.5 MeV wide $\{E_{20} \div E_o\}$, where $E_{20} = E_o - 3.5$ MeV.

Then one chose under what angle $\Theta_{Lab}$ to make measurements.

1) The exposition "0" starts from $E_o$ – the moderator in position "0" – open window.

Acquisition system starts measurement of first typical spectrum of particle-products with elastic and inelastic (if any) processes (Fig. 3 gray filled, or Fig. 4, group (R) spectrum with $A_o$, $B_o$, $C_o$, $H_o$). Measurement is done when the needed statistics under needed peak of process of interest ($A_o$, $B_o$, $C_o$, $H_o$) is accumulated.

IBM PC memory stores the measured spectrum with ($A_o$, $B_o$, $C_o$, $H_o$) and $Q_o$ total integrated beam charge.
                No stop of cyclotron !
2) Exposition "1" starts when moderator installs into beam thickness #1 plate. One will have beam-energy $E_1$ in a second (from Tab. 1) without touching whole beam-pipe optics or cyclotron.

Acquisition system starts measurement next typical spectrum of particle-products with elastic and inelastic (if any) processes (Fig. 4, group (R) spectrum with $A_1$, $B_1$, $C_1$, $H_1$. Measurement is done when the needed statistics under needed peak of process (of interest $A_1$, $B_1$, $C_1$, $H_1$) is accumulated in terms of $Q_1 = K_1 \times Q_o$.

Computers memory stores the measured spectrum with ($A_1$, $B_1$, $C_1$, $H_1$) and $Q_1$ total integrated beam charge…

20) Last exposition "19" results in products-spectrum with peaks $A_{19}$, $B_{19}$, $C_{19}$, $H_{19}$ and computers memory stores it together with $Q_{19} = K_{19} \times Q_o$.

First cycle done and one can repeat it if statistics or any other cause requires (something wrong was found…).

The fast TDC has time-resolution ~ 10 picoseconds, what allowed to use 1024 channels of coordinate scale with final resolution ≤0.2 mm along 300 mm of the "Apelsin" focal-plane (MWPC entrance window).

In the real experiment the 512-channel scale was used. Integral and differential nonlinearity of the coordinate scale finely were 0.4% and 1.6% correspondingly.

Tilt of the product-particle trajectory to the focal plane of "Apelsin" was ~ 36° (for the MWPC upper counter anode axis).

## 6. Data processing

Real beam-particle superposition spectrum (Q on Fig.4) is not flat and one have to make correction. That's why every peak $A_i$ from the $A_\Sigma$ was processed first by moving-average smoothing 11 times among 5 bins, then each resulted peak was fitted to a set of main $MG_i$ (central) + two (left and right side) background Gaussians (*MINUIT*).

The MWPC also has its own "non flat" registration effectivity – $CE_i$ along the anode wire axis. It was studied for protons, $\alpha$, $\beta$, and X-rays. The normalized $CE_i$ curve is - $NCE_i$ -it must be taken into account.

The result – the **normalized to 1** sum of all 20 smoothed main peaks, treated by the $NCE_i$ curve, gives the final correction curve "UMDCF".

Almost the same result is reachable if one applies the *Au* gold leaf target.

Such a special processing by 10-parameter *MINUIT* is applied because of the delta-electrons influence especially around spots of high-intensity (peaks on the focal plane – main anode of MWPC).

**Elastic EF extraction** was obtained on first step by subtracting $MG_i$ from each $A_i$ peak (Fig.4, (S)-group):

$$EF_i = A_i - MG_i \qquad , (8)$$

Then the correlation controlled oversewing (sum) of all 20 $EF_i$-peaks on one 512-bin scale gives the raff elastic **EF** which, after correction by UMDCF, becomes the final elastic $EF_i$ statistics.

The last one has to be transposed to the relativistic momentum scale by (9) with some of "Apelsin" and MWPC constants:

$$S[pc_i] = EF_i \frac{[80+1.25(X_i+C_{LCN}+CL_{CHI}+L_{VB}+Y/tg(36°))]}{A_{CMB} \, W_{oCN}} \, , (9)$$

Where $X_i$-is transposition of TDC-scale onto real focal plane, $A_{CMB}$ -is fraction of spectral line accepted by anode counter in the given section (#7) of spectrograph and $W_{oCN}$ – solid acceptance angle of this spectrograph analyzing sector (#7) in steradians.

Then $S(pc_i)$ was transposed to the projectile energy-scale, where the binning was calculated in decimal metric scale, what resulted in $S(E_j)$.

Real cross-section, as function of projectile-energy $E_j$, was calculated in [mb/sr] via (10):

$$\frac{d\sigma(E_j)}{d\Omega_{LAB}} = \frac{S(E_j) \, A}{h \, S_T \, N_A \left[ \frac{U_{sp}}{\Delta J} \frac{Q_{int}}{1.602 \cdot 10^{-19} Q} \frac{10^{-6} Q}{991} \right]} \, , \quad (10)$$

with $S(E_j)$ -the product-particle statistics on energy-scale, $A$- atomic weight of the target element, $h$ -target thickness [mg/cm$^2$], $S_T$ – area of the beam-spot on target [cm$^2$], $N_A$ - Avogadro's number, $U_{sp}$ -coefficient of used fraction of total spectrum of event statistics, $\Delta J=J_{max}-J_{min}$ - number of the used channels on spectrum, $Q_{int}$ -total number of beam-integrator pulses.

This processing finally results in a number of files:
- 1 -of reaction statistics on *pc*-scale of rel.momentum;
- 2 -of reaction statistics on scale of $E_{beamLAB}$;
- 3 file of $d\sigma(E_{beamLAB})/d\Omega_\theta$ for $\{E_{20} \div E_o\}$ energy range of 3.5 MeV.

## 7. Computer simulation

To ensure correctness of data processing a computer simulation was performed in which the artificial EF was first generated by means of URAND software. Then it was processed by convolutions to get 20 separate peaks with piloting displayed EF fractions on them.
After that whole MSS-algorithm was applied. The simulation showed stability of the results – the artificial EF was restored with errors less than 15% when the centroids of "beam" peaks where totally unknown or got shifted (magnetic field or accelerator's energy temporary change…).
In the case of the known medians of "beam"-peaks the residual errors were much smaller than 5%.

## 8. MSS -Excitation Function of $^{12}$C(p,p$_o$)

The *EF* of (*p*, $^{12}C$ ), measured in the energy range of (16 MeV÷19.5 MeV) by the MSS-approach, showed a resonance reach structure newer observed before in this energy region [1,4,14]. Comparison with [18] easily points on it (set of Δ on the Fig.7 below ). Some of peaks are well known and precisely fit the energy thresholds & levels from large amount of data [15-17].

The other peaks are unknown resonances in the excited system of 13 nucleons and the way of their "decoding" is really a wonder described in next preprint.

## Reading 0.

$$E_{pRF} = E^*_{LNP} + E^*_{RC} + \Delta M_{OS} - \Delta M_{PS}$$

- $E_{pRF}$ — proton rest frame energy
- $E^*_{LNP}$ — level energy of nucleus-product
- $\Delta M_{OS}$ — mass excess of output system
- $\Delta M_{PS}$ — mass excess of primary system

$\Delta M_{OS} = \Delta M_{NP} + \Delta m_{RC}$

$\Delta M_{PS} = \Delta M_{TN} + \Delta m_{BP}$

- $\Delta M_{NP}$ — mass excess of nucleus-product
- $\Delta M_{TN}$ — mass excess of target-nucleus
- $\Delta m_{RC}$ — mass excess of residual product-cluster(s) or/and particle(s)
- $\Delta m_{BP}$ — mass excess of beam-particle

**Writings I and II**

for all known product-nuclei in the **IS**

lead to the **Reading 0**

| | | |
|---|---|---|
| ← 7.2748 | ∬ 3 $^4He^*$ | (+ p) |
| ← 7.3665 | ∬ $^8Be^*$ | (+ $^4He$ + p) |
| ← 7.5516 | ∬ $^9B^*$ | (+ $^4He$) |
| ← 9.2388 | ∬ $^5Li^*$ | (+ 2 $^4He$) |
| ← 9.3306 | ∬ $^8Be^*$ | (+ $^5Li$) |
| ← 12.3706 | ∬ $^5Li^*$ | (+ $^8Be^{1*}$) |

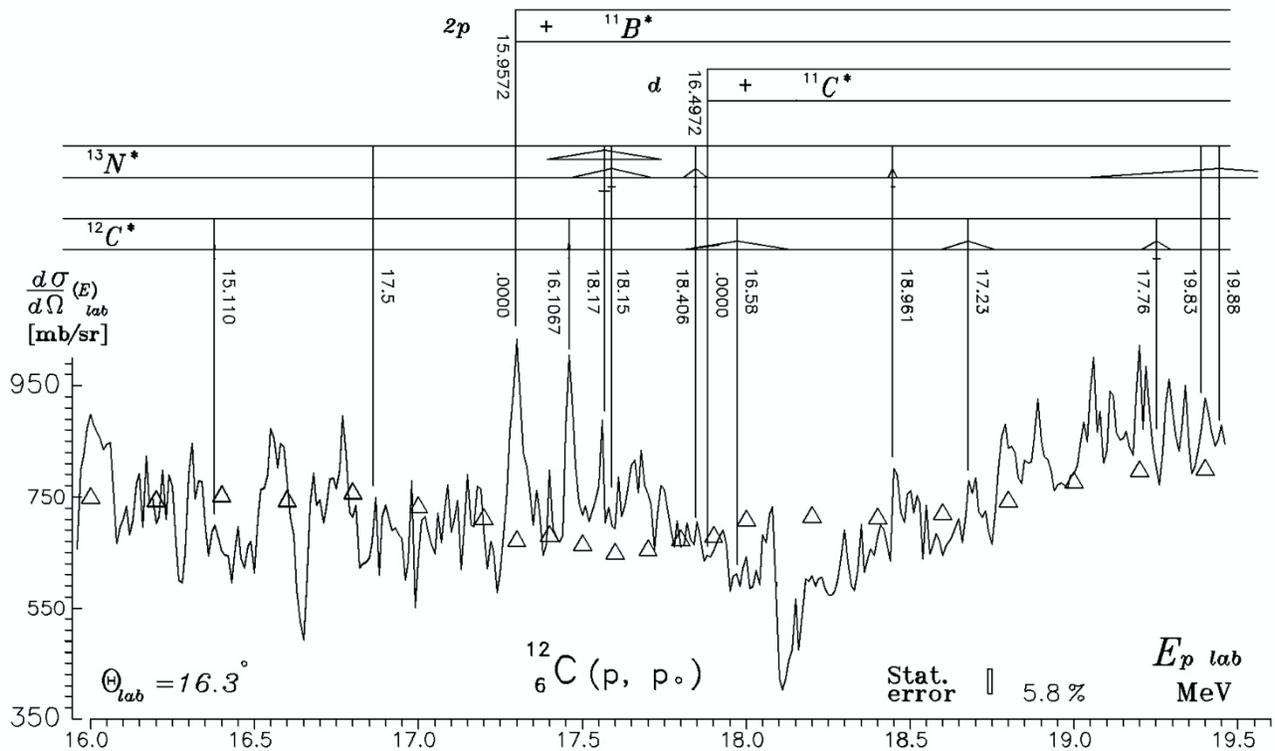

Fig. 7. Excitation function of $^{12}C(p, p_o)$ measured in the range 16 ÷ 19.5 MeV by MSS-approach at magnetic spectrograph "Apelsin" and U-150 cyclotron beam with $E_p$ =19.5 MeV with energy-spread ~200 keV.
Δ - from W. W. Daehnick and R. Sherr. Phys. Rev. 133, B934

## 9. Conclusion

The MSS-principle is real, and it works well!

It is very important to stress that the elastic scattering, for the forward angles, has largest cross-section among all other processes.

On the other hand, elastic scattering **carries all resonances** (traces) of all nuclear cluster-combinations (possible for given (target + projectile)-system) that are reachable for the given projectile energy.

Most of these cluster-combination do not survive to the "finish" being "executed" by the interior decay due to conservation laws and quantum boundaries.

But (!) formation (population) of these cluster-combination is directly visible in the cross-section of elastic scattering, which is like a magnifying glass, - shows all the "subtitles" of internuclear processes!

Such a "magnifying glass" allows to bring a new breading to modern nuclear physics especially at the RIBs and exotic nuclear targets, simultaneously providing precise information due to MSS, for example.

Next preprint would be devoted to deep description of the measured excitation function and "decoding" of the displayed unknown resonance anomalies.

## 10. Acknowledgements

I would like to thank scientist who started this branch of experimental approach.

They invested and pushed us to make it real as the MSS! Academician Bekhzad Yuldashev and professor Yury Koblik - my scientific gurus, also professor Vladimir Nikitin (JINR), Dr. Dmitry Mirkarimov, professor, Dr. B.S. Mazitov, Dr. V. Sidorov, Dr. V. Ulanov, Dr. S. Artiomov, Dr. Avas Khugaev, engineers V. Iakushev, A. Kadishnov, O. Sazonov, A. Tsupin, I. Atabaev.

Special thanks to my official opponents professor Vladimir Tokarevsky (KINR) and Boris Skorodumov.

Big thanks to Dr. M. Golovkov (Kurchatov Institute).

My sincere gratitude to BU-scientists professor James P. Miller and professor Robert M. Carey.